\documentstyle[12pt]{article}
\def\nab#1{{\nabla_{#1}}}
\def\nabstar#1{\nabla\kern-0.5pt\smash{\raise 4.5pt\hbox{$\ast$}}
               \kern-4.5pt_{#1}}

\def\drvstar#1{\partial\kern-0.5pt\smash{\raise 4.5pt\hbox{$\ast$}}
               \kern-5.0pt_{#1}}

\def\newline{\relax\ifhmode\null\hfil\break\else\nonhmodeerr@\newline\fi}
\def\frac#1#2{{#1\over#2}}
\def\text#1{{\hbox{\rm #1}}}
\def\flushpar{{\par \noindent}}

\newcommand{\beq}{\begin{equation}}
\newcommand{\eeq}{\end{equation}}
\newcommand{\bea}{\begin{eqnarray}}
\newcommand{\eea}{\end{eqnarray}}
\def\Id{ \mbox{1\hspace{-1.2mm}I} }
\def\BE{\begin{equation}}
\def\EE{\end{equation}}
\def\BA{\begin{eqnarray}}
\def\EA{\end{eqnarray}}
\def\BAN{\begin{eqnarray*}}
\def\EAN{\end{eqnarray*}}

\def\nn{\nonumber\\}

\def\tr{\mbox{tr}}

\def\gm5{\gamma^5}

\input epsf.sty
\newdimen\psfigsize
\def\psfigure#1 #2 #3 #4 #5{
    \begin{figure}[tbh]
      \begin{center}
      \vbox{
        \null\vskip-0.2in\hskip#2
        \epsfxsize=#1
        \epsfbox{#4}
        \vskip -0.3in
        \caption {#5 \label{#3}}
        \vskip 0.0 true in plus 0.3 true in
      }
      \end{center}
   \end{figure}
}
\begin{document}
\thispagestyle{empty}
\begin{flushright}
NTUTH-98-097 \\
September 1998
\end{flushright}
\bigskip\bigskip\bigskip
\vskip 2.5truecm
\begin{center}
{\LARGE {Topological phases in Neuberger-Dirac operator}}
\end{center}
\vskip 1.0truecm
\centerline{Ting-Wai Chiu}
\vskip5mm
\centerline{Department of Physics, National Taiwan University}
\centerline{Taipei, Taiwan 106, Republic of China. }
\centerline{\it E-mail : twchiu@phys.ntu.edu.tw}
\vskip 2cm
\bigskip \nopagebreak \begin{abstract}
\noindent

The response of the Neuberger-Dirac fermion operator $ D = \Id + V $ in
the topologically nontrivial background gauge field depends on the
negative mass parameter $ m_0 $ in the Wilson-Dirac fermion operator
$ D_w $ which enters $ D $ through the unitary operator
$ V = D_w ( D_w^{\dagger} D_w )^{-1/2} $.
We classify the topological phases of $ D $ by comparing
its index to the topological charge of the smooth background gauge field.
An exact discrete symmetry in the topological phase diagram is proved for
any gauge configurations. A formula for the index of $ D $ in each
topological phase is derived by obtaining the total chiral charge of
the zero modes in the exact solution of the free fermion propagator.

\vskip 2cm
\noindent PACS numbers: 11.15.Ha, 11.30.Fs, 11.30.Rd

\end{abstract}
\vskip 1.5cm

\newpage\setcounter{page}1

\section{ Introduction }

It is well known that the Neuberger-Dirac operator \cite{hn97:7,hn98:2}
\bea
\label{eq:Dh}
D = \Id + V, \hspace{4mm} V = D_w ( D_w^{\dagger} D_w )^{-1/2}
\eea
reproduces the exact zero modes with definite chirality and realizes the
Atiyah-Singer index theorem on a finite lattice for smooth background
gauge fields with nonzero topological charge \cite{twc98:4}.
However, this is only true for the negative mass parameter $ m_0 $
approximately in the range $ ( 0, 2 r_w ) $ for the Wilson-Dirac fermion
operator $ D_w $ with Wilson parameter $ r_w > 0 $
\bea
\label{eq:Dw}
D_w = - m_0
      + \frac{1}{2} [ \gamma_{\mu} ( \nabstar{\mu} + \nab{\mu} ) -
                      r_w  \nabstar{\mu} \nab{\mu} ]
\eea
where $ \nab{\mu} $ and $ \nabstar{\mu} $ are the forward and
backward difference operators defined in the following,
\bea
\nab{\mu}\psi(x) &=& \
   U_\mu(x)\psi(x+\hat{\mu})-\psi(x)  \nonumber \\
\nabstar{\mu} \psi(x) &=& \ \psi(x) -
   U_\mu^{\dagger}(x-\hat{\mu}) \psi(x-\hat{\mu})  \nonumber
\eea
Except for the region $ m_0 < 0 $ which corresponds to the positive mass in
$ D_w $ that gives zero index of $ D $ for any background gauge field has been
noted by Neuberger in ref. \cite{hn98:2}, there is literally no discussions
in the literature about what would happen to $ D $ for other values of
$ m_0 $, namely, $ m_0 \ge 2 r_w $. In ref. \cite{chand98:5}, it is
remarked that $ D_w $ must have an appropriately tuned negative mass term
such that $ D $ is in the right phase to reproduce QCD. However, there is no
further discourse on how to identify the phase such that $ D $ can possess
the proper topological characteristics and reproduce the correct continuum
physics.

In this paper we investigate the topological phases of $ D $ with respect
to the negative mass parameter $ m_0 $ and the background gauge field
respectively, by comparing its $ \mbox{index}(D) $
to the topological charge $ Q $ of the background gauge field.
If $ \mbox{index}(D) = 0 $ for any
smooth gauge configurations, then $ D $ is called to be in the
{\it topologically trivial phase.} If $ \mbox{index}(D) = Q $,
then $ D $ is called to be in the {\it topologically proper phase.}
If $ D $ is not topologically trivial but $ \mbox{index}(D) \ne Q $,
then $ D $ is called to be in the {\it topologically improper phase.}

First, we sketch schematically the topological phases of $ D $ in
Fig. \ref{fig:dd}. On a $d$-dimensional lattice\footnote{In this paper
we always assume that $d$ is a positive even integer.},
the topological phases of $ D $ in the free fermion limit divide $ m_0 $
values into $ d + 2 $ intervals according to the real eigenmodes of $ D $.
As we turn on the background gauge field, the locations of the phase
boundaries also evolve accordingly. The first region cuts the interval
$ m_0 < 0 $ in the free fermion limit.
In this phase, the spectrum of $ D $ does not possess
any zero modes and thus cannot describe massless particles, and
the index is zero for any background gauge field. Therefore $ D $ is called
{\it topologically trivial } in this phase. The second region cuts
the interval $ 0 < m_0 < 2 r_w $ in the free fermion limit. In this phase,
$ D $ has exact zero modes with definite chirality, and the index of $ D $
is equal to the topological charge $ Q_{top} $ of the background gauge field.
Therefore $ D $ is called {\it topologically proper} in this phase.
The third region cuts the interval $ 2 r_w < m_0 < 4 r_w $ in the free
field limit. In this phase, $ D $ also has exact zero modes with definite
chirality, {\it but the index of $ D $ is $ (1-d) Q_{top} $, not equal to
the topological charge of the background gauge field.}
Thus $ D $ is called {\it topologically improper } in this phase.
The next phase again has width $ 2 r_w $ in the free fermion limit.
The process of dividing $ m_0 $ into different topological phases
continues until it reaches $ m_0 = 2 d r_w $ in the free fermion limit,
then the last phase is the region with the edge, $  2 d r_w <  m_0 < \infty $.
In any dimensions, only the region cutting the interval $ 0 < m_0 < 2 r_w $
in the free fermion limit is the topologically proper phase for $ D $,
other phases are either topologically trivial or improper.
The index along each one of the phase boundaries starting at
( $ m_0 = 2 k r_w, k=0,\cdots,d $ ) is zero identically for
any background gauge field.

In the free fermion limit, we can distinguish different
phases of $ D $ according to its real eigenmodes. {\em The phase
transition is signaled by the change of the numbers of real eigenmodes.}
We note that in the free fermion limit the real eigenmodes of $ D $ must be
either at the origin or the corners of the Brillouin zone. However, the
converse of this statement is not always true, since {\em at some values of
$ m_0 $, some of the eigenmodes at the corners of Brillouin zone become
complex}. Then a formula for the index of $ D $ in each phase can be derived
( in Section 3 ) by obtaining the total chiral charge of the zero modes in
the exact solution of the free fermion propagator.
For smooth background gauge fields, this formula (\ref{eq:index_pert})
gives the correct index in each phase except at the phase
boundaries ( $ m_0 = 2 k r_w, k=1,\cdots,d $ ).

For any arbitrary gauge configuration, we prove that there exists an exact
reflection symmetry in the topological phase diagram with
respect to the symmetry axis $ m_0 = d r_w $, i.e.,
$ \mbox{index}[ D(m_0) ] = - \mbox{index} [ D( 2 d r_w - m_0 )] $,
for a finite lattice of even number of sites in each dimension.
It follows that if there is a phase boundary located
at $ m_0 = x < d r_w $ on the LHS, then there is another phase boundary
located at $ m_0 = 2 d r_w - x $ on the RHS, and these two phase boundaries
appear as images to each other with respect to the mirror at $ m_0 = d r_w $.
Since $ d $ is an even integer, the exact reflection symmetry implies that
{\em the symmetry axis $ m_0 = d r_w $ is a phase boundary with zero index
and its location is invariant for any background gauge configurations.}
( for example, see Fig. 2 and Fig. 3 )

As we turn on the background gauge field smoothly from
zero to some finite values, {\em the locations of the phase boundaries also
evolve accordingly}. When the gauge fields reach a certain level of
magnitude and roughness, {\em the phase boundaries bifurcate} and the
phase diagram becomes very complicated.
We expect that for gauge configurations in dynamical simulations
the phase boundaries may change significantly from those of smooth
background gauge fields. However, the exact reflection symmetry
must hold in general and provides a useful scheme to identify the
topological phases especially for rough gauge configurations.
We note that at one of the topological phase boundaries of $ D $,
the theory is free of species doublings in the free fermion limit but
turns out to be topologically trivial in any background gauge fields.
This provides an explicit example that for $ D $ free of species
doubling in the free fermion limit is not sufficient to gaurantee the
realization of Atiyah-Singer index theorem on the lattice, in contrast to
the condition of reproducing correct axial anomaly in perturbation
calculations \cite{twc99:1}. This may indicate that the topological
characteristics of the fermion operator $ D $ are of nonperturbative
origins and thus could not be revealed by any perturbation calculations.
We note in passing that the locality of Ginsparg-Wilson fermion operator
$ D $ is not relevant to its zero modes and index, as discussed and
demonstrated in ref. \cite{twc98:9a,twc98:9b}.

\psfigure 5.0in -0.2in {fig:dd} {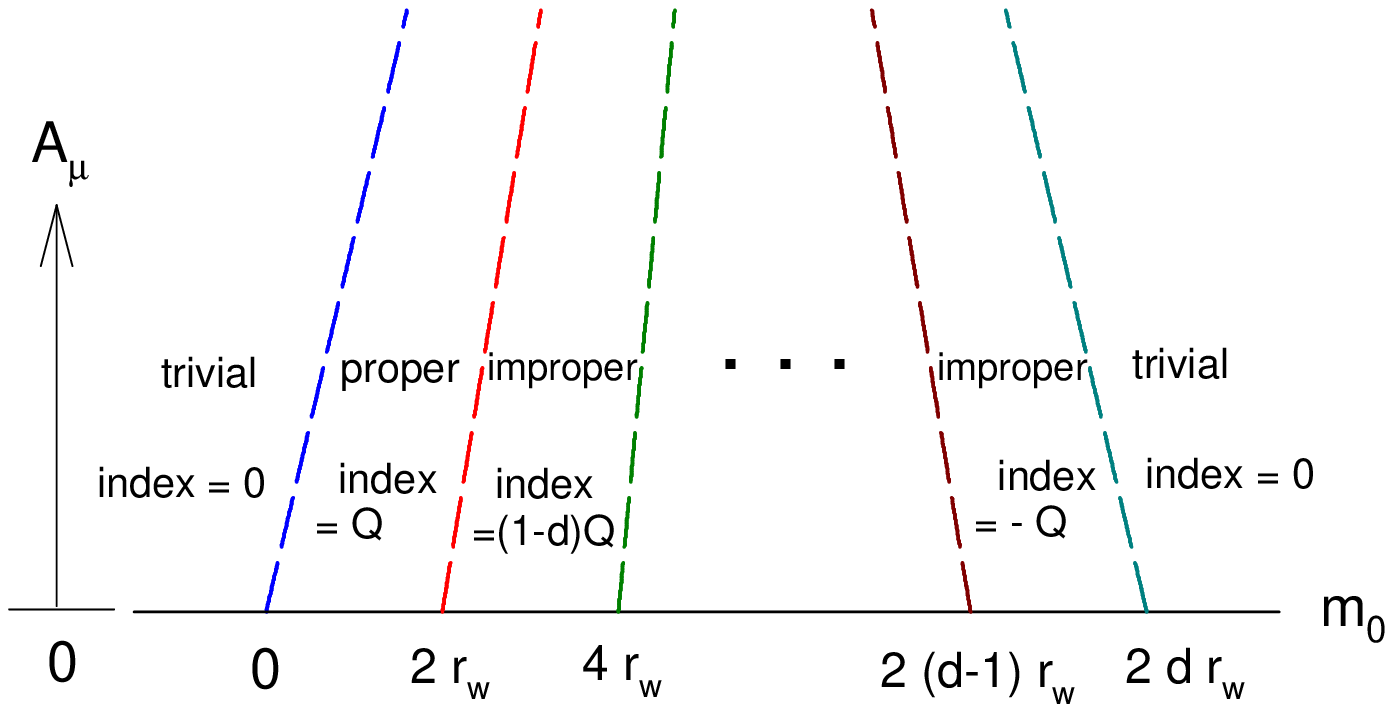} {
The topological phase diagram of $ D $ on a $d$-dimensional lattice
for smooth background gauge field.
The phase boundaries denoted by dashed lines are sketched to indicate
their evolutions with respect to the background gauge field.
The critical points in the free fermion limit ( at the horizontal line )
are determined exactly. The reflection symmetry with respect to the
symmetry axis $ m_0 = d r_w $ is also shown explicitly. }

The outline of this paper is as follows. In Section 2, we prove the
exact reflection symmetry in the topological phase diagram of $ D $.
In Section 3, we obtain the exact solutions of the eigenvalues and the
fermion propagator of $ D $ in the free fermion limit, then use them to
analyze the topological phases of $ D $. A formula for the index of $ D $
in each phase is derived. In Section 4, we perform numerical experiments
to investiagate the topological phase diagrams of $ D $ in two dimensional
and four dimensional lattices respectively. In Section 5,
we summarize, and also demonstrate the bifurcation of topological
phases by increasing the amplitudes of the local sinusoidal fluctuations
of the background gauge field on a two dimensional lattice. In the Appendix,
we discuss the spectral flow of the Neuberger operator as a function
of the negative mass parameter $ m_0 $, in particular, its behaviors
at the phase boundaries, in order to gain a perspective on the underlying
mechanism which gives the zero index at the phase boundaries.

\section{ The discrete symmetry of the index }

In this section, we prove that for a finite lattice with {\it even}
number of sites in each dimension
( $ L_\mu = N_\mu a, N_\mu = \mbox{even integer}, \mu = 1, \cdots, d $ ),
the topological phase diagram of the Neuberger-Dirac operator $ D(m_0) $
has an {\it exact } reflection symmetry with respect to the axis
$ m_0 = d r_w $, and the index of $ D $ is {\it odd} under the reflection,
i.e., $\mbox{index}[ D(m_0) ]=-\mbox{index} [ D( 2 d r_w - m_0 )] $. \newline
{\it Proof : } \newline
For a given gauge configuration, the index of $ D(m_0) $ is defined to
be $ n_{-}(m_0) - n_{+}(m_0) $, where $ n_{-} ( n_{+} ) $ denotes
the number of zero modes of negative ( positive ) chirality.
Let $ \phi_{+}^s, s=1, \cdots, n_{+} $ be the normalized eigenfunctions of
the zero modes with positive chirality,
\bea
D(m_0) \phi_{+}^s = 0, \hspace{4mm} \gamma_5 \phi_{+}^s = \phi_{+}^s.
\eea
Then $ \phi_{+}^s $ is a real eigenmode of $ D_w $ with negative real
eigenvalue $ -\mu_s $ such that
\bea
V(m_0) \phi_{+}^s &=& - \phi_{+}^s \\
D_w(m_0 ) \phi_{+}^s &=& -\mu_s \phi_{+}^s
\label{eq:Dw_eig}
\eea
Similar equations for zero modes with negative chirality are understood.
The standard Wilson-Dirac lattice fermion operator $ D_w $,
Eq. (\ref{eq:Dw}) can be rewritten as
\beq
\label{eq:DwT}
[ D_w ( m_0 ) ]_{x,y} = ( d r_w - m_0 ) \delta_{x,y} - [T(r_w)]_{x,y}
\eeq
where
\beq
\label{eq:T}
[T(r_w)]_{x,y} = \frac{1}{2} \sum_{\mu} \left[
   ( r_w - \gamma_{\mu} ) U_{\mu}(x) \delta_{x+\hat\mu,y}
 + ( r_w + \gamma_{\mu} ) U_{\mu}^{\dagger}(x-\hat\mu) \delta_{x-\hat\mu,y}
 \right]
\eeq
The Dirac and color indices are suppressed. The fundamental properties of
$ T $ are \cite{gatt97:7} :
\bea
\label{eq:herm}
\gamma_5 T \gamma_5 = T^{\dagger} \\
\label{eq:ST}
S T S = - T
\eea
where $ S_{x,y} = \prod_{\mu} (-1)^{x_\mu/a} \delta_{x,y} $ which satisfies
$ S^2 = \Id $ and $ S^{\dagger} = S $. The first property is the usual
hermiticity condition. The second property is also obvious except at the
boundaries of the lattice ( with periodic boundary conditions ), where the
forward and the backward difference operators in $ T $ would produce a factor
+1 ( -1 ) for odd ( even ) number of sites in this direction. Therefore
Eq. (\ref{eq:ST}) is satisfied exactly on a finite lattice having even number
of sites in each dimension, but only satisfied approximately on a finite
lattice having odd number of sites in some dimensions. Of course, it is
satisfied exactly on the infinite lattice.
It is evident that an eigenfunction of $ D_w $ must be
an eigenfunction of $ T $, and vice versa. Then from Eqs. (\ref{eq:Dw_eig})
and (\ref{eq:DwT}), we have
\bea
T \phi_{+}^s = t_s \phi_{+}^s
\eea
where $ t_s = \mu_s + ( d r_w - m_0 ) $. Using Eq. (\ref{eq:ST}) and
$ S^2 = \Id $, we obtain
\bea
  \mbox{det}( T - t \Id )
&=& \mbox{det}(S) \ \mbox{det}( T - t \Id ) \ \mbox{det}(S) \nn
&=& \mbox{det}( S T S - t \Id ) = \mbox{det}( - T - t \Id ) \nonumber
\eea
Thus the eigenvalues of $ T $ must come in $ \pm t_s $ pairs. It follows
that the eigenvalues of $ D_w(m_0) $ must come in $ d r_w - m_0 \pm t_s $
pairs. Now if we substitute $ m_0 $ by $ 2 d r_w - m_0 $, then the
eigenvalues of $ D_w( 2 d r_w - m_0 ) $ come in $ -( d r_w - m_0 \mp t_s ) $
pairs. Therefore, from Eq. (\ref{eq:Dw_eig}), if $ D_w(m_0) $ has a negative
real eigenvalue $ -\mu_s $ for the eigenfunction $ \phi_{+}^{s} $, then
$ D_w( 2 d r_w - m_0 ) $ has a corresponding positive real eigenvalue,
$ +\mu_s $ with the same eigenfunction $ \phi_{+}^{s} $, and this implies
that $ D( 2 d r_w - m_0 ) $ has a $ +2 $ real eigenmode while
$ D(m_0) $ has a zero mode, and vice versa, i.e.,
\bea
\label{eq:n2+}
n_2^{+} ( m_0 ) = n_{+} ( 2 d r_w - m_0 ) \\
\label{eq:n2-}
n_2^{-} ( m_0 ) = n_{-} ( 2 d r_w - m_0 ) \\
\label{eq:n+}
n_{+} ( m_0 ) = n_2^{+} ( 2 d r_w - m_0 ) \\
\label{eq:n-}
n_{-} ( m_0 ) = n_2^{-} ( 2 d r_w - m_0 )
\eea
As it was first proved in ref. \cite{twc98:4} that the number of zero
modes and the number of $ +2 $ modes for $ D(m_0) $ must obey the relation
\bea
\label{eq:realmodes}
n_2^{+}(m_0) - n_2^{-}(m_0) = n_{-}(m_0) - n_{+}(m_0),
\eea
we immediately obtain
\bea
n_{+} ( 2 d r_w - m_0 ) - n_{-}( 2 d r_w - m_0 ) = n_{-}(m_0) - n_{+}(m_0).
\eea
That is
\bea
\label{eq:ref_sym_1}
\mbox{index}[ D( 2 d r_w - m_0 ) ] = - \mbox{index} [ D( m_0 )].
\eea
{\it This is the discrete symmetry of $ \mbox{index}(D) $ which holds for
any gauge configuration.}
It transcribes to the reflection symmetry in the topological phase
diagram with the symmetry axis at $ m_0 = d r_w $. When the gauge
fields are very rough, the topological phase diagram becomes very
complicated, the only thing surely survives in such chaotic
environments is the exact discrete symmetry, Eq. (\ref{eq:ref_sym_1}).
For a finite lattice having {\it odd} number of sites in some directions,
the reflection symmetry Eq. (\ref{eq:ref_sym_1}) is only an approximate
discrete symmetry which tends to the exact reflection symmetry in the
infinite lattice limit. One immediate consequence of the exact discrete
symmetry (\ref{eq:ref_sym_1}) is that at the symmetry axis
$ m_0 = d r_w $, the index is zero for any background gauge field,
\bea
\label{eq:sym_axis}
\mbox{index}[ D( d r_w ) ] = 0
\eea
Since $ d $ is an even integer,
{\em the symmetry axis $ m_0 = d r_w $ is a phase boundary with zero index
and its location is invariant for any background gauge configurations.}
( for example, see Fig. 2 and Fig. 3 ).

We also note that $ T(r_w) $ has another discrete symmetry,
\bea
[ T(r_w) ]^{\dagger} = - T(-r_w)
\eea
which follows directly from Eq. (\ref{eq:T}). Then we have the following
discrete symmetry for $ D_w $,
\bea
[ D_w ( m_0, r_w ) ]^{\dagger} = - D_w ( -m_0, -r_w )
\eea
From this relation, we immediately know that the zero modes of $ D(m_0,r_w) $
are the $ +2 $ eigenmodes of $ D( -m_0, -r_w ) $, and vice versa.
Using Eq. (\ref{eq:realmodes}), we obtain the following discrete symmetry for
the index of $ D $
\bea
\label{eq:ref_sym_2}
\mbox{index}[ D( m_0, r_w ) ] = - \mbox{index} [ D( -m_0, -r_w )].
\eea
This discrete symmetry implies that if we change the sign of the Wilson
parameter $ r_w $, then the new topological phase diagram is the
{\it negative} image of the old topological phase diagram with respect
to the mirror at $ m_0 = 0 $. This reflection symmetry will not be
exploited in this paper.

We remark that the discrete symmetries,
Eqs. (\ref{eq:ref_sym_1}) and (\ref{eq:ref_sym_2}) discussed above also
hold for the generalized Neuberger-Dirac operator \cite{twc98:9a}
\bea
\label{eq:DV}
D = 2 M ( \Id + V ) [ ( \Id - V ) + 2 M R ( \Id + V ) ]^{-1}
\eea
where $ R $ is a non-singular hermitian local operator
which is trivial in the Dirac space,
$ M $ is an arbitrary mass scale, and $ V ( m_0, r_w ) $ is
the unitary operator defined in Eq. (\ref{eq:Dh}). The reason for these
discrete symmetries holding for the Ginsparg-Wilson generalized $D$ in
(\ref{eq:DV}) is that the zero modes of $ D $ are invariant for any
$ R $ \cite{twc98:9a,twc98:9b}, and hence any relationships involving only
the indices must be $R$-invariant. The proof for the general case is obvious
from the above proof.

\section{The exact solutions in the free fermion limit}

In the free fermion limit, the index of $ D $ must be zero,
but we can use the number of zero modes of $ D $ to distinguish
different topological phases of $ D $. Since there is a symmetry between
the number of zero modes and the number of $+2$ modes as first shown in
ref. \cite{twc98:4}, Eqs. (\ref{eq:n2+})-(\ref{eq:n-}), then it suffices
to use both the number of zero modes and the number of $+2$ eigenmodes as
the order parameter for the topological phases of $ D $ in the free fermion
limit. Then a phase transition is signaled by any changes of these two numbers.
Alternatively, we can examine the fermion propagator and use the number
of massless ( primary and doubled ) fermion modes it contains to distinguish
different topological phases of $D$. From the total chiral charge $ Q_5 $
of the massless fermion modes in the fermion propagator, we can determine
the integer multiplier of the $ F \tilde{F} $ term in the axial anomaly
$ q(x;A,D) = tr[ \gamma_5 (R D) (x,x) ] $ for $ D $ in a smooth background
gauge field of topological charge $ Q_{top} $, which also can be determined
by explicit perturbation calculations. Furthermore we can also predict that
the index of $ D $ is equal to the total chiral charge times the topological
charge of the background gauge field, i.e.,
\bea
\label{eq:indQ5}
\mbox{index}(D) = Q_5 Q_{top}
\eea
This expression is deduced straightforwardly from the weak coupling
perturbation theory, which in fact correctly predicts the index of $ D $ in
smooth background gauge fields for all values of $ m_0 $ except at the
phase boundaries starting at $ m_0 = 2 r_w,  4 r_w, \cdots, 2d r_w $,
where the index is always zero regardless of the total chiral charge of
the zero modes. The disagreement between the prediction (\ref{eq:indQ5})
and the zero index at the phase boundaries, indicates that there may be
some nonperturbative or topological effects which cannot be taken into
account in the perturbation theory. We will show that
Eq. (\ref{eq:indQ5}) violates the exact reflection symmetry at the phase
boundaries.

First, we obtain the exact solutions of the eigenvalues and the
fermion propagator of $ D $.
In the free fermion limit, the eigenvalues of the Neuberger-Dirac
operator $ D $ [ Eq. (\ref{eq:Dh}) ] on a $d$-dimensional lattice
[ $ L_\mu = N_\mu a, \ \mu=1, \cdots, d ( \mbox{even integer} ) $ ] with
periodic boundary conditions are
\bea
\label{eq:eigen}
\lambda_s = 1 + \frac{u(p)}{N(p)}
            \pm i \frac{ \sqrt{ \sum_{\mu=1}^{d} \sin^2(p_\mu a ) } }{N(p)}
\eea
where
\bea
u(p)  &=& - m_0 + r_w \sum_{\mu=1}^{d} \left[ 1 - \cos(p_\mu a) \right]    \\
N(p)  &=& \sqrt{ u^2(p) + \sum_{\mu=1}^{d} \sin^2(p_\mu a) }
\eea
\beq
p_\mu  = \left\{  \begin{array}{ll}
\frac{-\pi(N_\mu-2)}{L_\mu}, \cdots, \frac{\pi(N_\mu-2)}{L_\mu},
\frac{\pi}{a}                   &      \mbox{ if $ N_\mu $ is even }  \\
\frac{-\pi(N_\mu-1)}{L_\mu}, \cdots, \frac{\pi(N_\mu-3)}{L_\mu},
\frac{ \pi(N_\mu-1)}{L_\mu}     &      \mbox{ if $ N_\mu $ is odd  }  \\
                  \end{array}
              \right.
\eeq
and each real $ \lambda_s $ has degeneracy $ 2^{[\frac{d}{2}]} $
consisting of chirality $ \pm 1 $ pairs in the Dirac space,
while each complex $ \lambda_s $ has degeneracy $ 2^{[\frac{d-2}{2}]} $.

The free fermion propagator in momentum space is \cite{twc98:4}
\bea
\label{eq:SF0}
S_F^{(0)}(p) = \frac{a}{2} \Id
  + \frac{a}{i \gamma_\mu \sin( p_\mu a ) } \ T(p)
\eea
where
\bea
\label{eq:Tp}
T(p) = \frac{1}{2}[ N(p) - u(p) ]
\eea
The constant term $ a/2 $ which vanishes in the continuum limit
can be ignored since its role is the same as the chirality breaking
operator $ R $ in the Ginsparg-Wilson fermion propagator. The second term
is equal to the naive lattice fermion propagator times the factor
$ T(p) $. The suppression of the doubled modes is due to the last
factor $ T(p) $ which becomes zero for some or all of the doubled modes
depending on the value of $ m_0 $. We have deliberately written the fermion
propagator in the form of Eq. (\ref{eq:SF0}) such that the decoupling
mechanism for the doubled modes is the most appealing to us.
It is interesting to note that the way doubled modes are decoupled in
(\ref{eq:SF0}) is very different from that
of the Wilson fermion propagator. The former is a complete suppression
even at finite lattice spacing, e.g., $ T(p) = 0 $ for all doubled
modes for $ 0 < m_0 \le 2 r_w $, while the latter is of order $ a $
which is completely decoupled only in the continuum limit ( $ a \to 0 $ ).
This implies that, {\em the Neuberger-Dirac fermion can reproduce the correct
continuum axial anomaly and the exact Atiyah-Singer index theorem at
finite lattice spacing, while the Wilson fermion can only recover them
in the continuum limit.} In other words, any ( primary or doubled )
massless fermion mode in the Neuberger-Dirac fermion propagator emerges
in a very clean-cut way, i.e., either existing or vanishing,
for any finite lattice spacing.

In order to discuss the chiral charge of the doubled modes, we consider
the free fermion propagator in the continuum limit ( $ a \to 0 $ ) on
the infinite lattice with the momentum space, the Brillouin zone ( BZ )
which is specified by
$ p_\mu \in ( -\pi/a, \pi/a ] $ for $ \mu = 1, \cdots, d $.
In the vicinity of the origin $p=0$, we can expand
$ \sin ( p_\mu a ) = p_\mu a + O( a^2 p_\mu^2 ) $ and
$ T(p) = m_0 + O ( a p^2 ) $ in Eq. (\ref{eq:SF0}) to obtain
\bea
\label{eq:SF00}
S_F^{(0)}(p) \stackrel{a,p \to 0}{ \longrightarrow }
- \frac{ i \gamma_\mu p_\mu }{p^2} m_0  + O(a^2 p )
\eea
where the constant term $ a/2 $ has been subtracted and
the factor $ m_0 $ can be absorbed by using the generalized
Dirac operator (\ref{eq:DV}) with $ M = m_0 $ and $ R = 1/2m_0 $,
thus $ D = m_0 ( \Id + V ) $.
Then $ S_F^{(0)}(p) $ in (\ref{eq:SF00}) agrees with the free fermion
propagator in continuum. This is the primary mode of the lattice fermion.
Besides the primary mode, it is also possible for $ S_F^{(0)}(p) $ to
develop massless doubled modes at other regions of the BZ due to the presence
of $ \sin(p_\mu a ) $ rather than $ p_\mu a $ in Eq. (\ref{eq:SF0}).
It is evident that the massless doubled modes can only occur around the
$ 2^d -1 $ corners of the BZ, i.e.,
$ \otimes_{\mu=1}^{d} \{ 0, \pi/a \} - \{ p=0 \} $.
If the factor $ T(p) $ vanishes smoothly at any one of these
corners, then the doubled mode at this corner is decoupled from
the theory, otherwise this doubled mode will contribute to the propagator.
Furthermore, a doubled mode may possess opposite chiral charge of
the primary mode due to the property of $ \sin( p_\mu a ) $.
When $ p_\mu $ is near $ \pi/a $, we must shift the momentum $ p_\mu $
to $ p_\mu^{'} = p_\mu - \pi/a $ such that the higher order terms
$ O(p^2 a^2) $ in $ \sin( p_\mu^{'} a ) $ can be neglected in the
continuum limit $ a \to 0 $. Then the relationship
$ \sin(p_\mu a) = - \sin(p_\mu^{'} a) $ produces an extra minus sign
which can be absorbed by redefining $ \gamma_\mu^{'} = - \gamma_\mu $.
Since $ \gamma_5 = \prod_{\mu=1}^{d} \gamma_\mu $, the effective
$ \gamma_5^{'} $ for a doubled mode is equal to $+\gamma_5 $ or
$-\gamma_5 $ depending on the number of $\pi/a$ momentum components
is even or odd. Since the sign of $ \gamma_5 $ is usually taken
to be the sign of the chiral charge, then we have $ 2^{d-1} - 1 $ doubled
modes of positive chiral charge, but $ 2^{d-1} $ of negative chiral charge.
After taking into account of the primary mode, the number of fermion
modes of positive chiral charge is equal to the number of negative ones.
If all these modes contribute to the propagator, then the total chiral
charge vanishes, the axial anomaly cancels, and the index is zero in
any background gauge field. That is the case for the naive lattice fermion.

For the Neuberger operator, the suppression of doubled modes depends
on the factor $ T(p) $ which is a function of $ m_0 $.
At the origin and the corners of the BZ, $ \sin( p_\mu a ) = 0 $,
then $ T(p) $ can be simplified to the following,
\bea
\label{eq:TpBZ}
T(p) = \frac{1}{2}[ | u(p) | - u(p) ]
\eea
where the possible values of $ u(p) $ are the following
\bea
\label{eq:uBZ}
u(p) = -m_0, -m_0 + 2r_w, -m_0 + 4r_w, \cdots, -m_0 + 2dr_w \
\eea
for these $ 2^d $ momenta. From Eq. (\ref{eq:TpBZ}), it is evident
that $ T(p) = 0 $ for $ u(p) \ge 0 $.

For $ m_0 \le 0 $, $ T(p) = 0 $ for all $ 2^d $ massless fermion
modes, hence they are all suppressed and decoupled from the theory.
Since the theory does not have any massless fermion modes, so the
index must be zero after the background gauge field is turned on.
This is a topologically trivial phase of $ D $.
Note that for $ m_0 \le 0 $, $ D $ can be shown to be topologically
trivial for any background gauge field since $ D_w $ does not possess
any negative eigenvalue for a positive mass parameter.

For all values of $ m_0 > 0 $, the primary mode at $ p=0 $ is recovered
to the theory since $ u(0) < 0 $ and $ T(0) \ne 0 $.
For $ 0 < m_0 < 2 r_w $, $ u(p) > 0 $ and $ T(p) = 0 $ for all
doubled modes, so all doubled modes are still decoupled from the theory.
When we turn on the background gauge field with topological charge $ Q_{top} $,
the primary mode then gives the index equal to $ +Q_{top} $, same as the
continuum theory. This is the topologically proper phase of $ D $.

At the phase boundary $ m_0 = 2 r_w $, all doubled modes have $ T(p) = 0 $,
hence they are decoupled from the theory. However, when we turn on the
background gauge field with topological charge $ Q_{top} $, the index is
always zero. From the viewpoint of perturbation theory, it is not clear what
has happened. If we examine the doubled modes more closely, we find that
there are $ d $ doubled modes which are just "critically" decoupled, i.e.,
each having only one nonzero ( $ \pi/a $ ) momentum component which gives
$ u(p) = 0 $ and $ T(p) = 0 $, hence they are decoupled from the theory.
On the other hand, if we assume that these $ d $ doubled modes still remain
in the theory, then the index should be $ (1-d) Q_{top} $ rather than zero,
since they carry negative chiral charge. So, this assumption cannot be true.
Moreover, $ T(p) $ goes to zero smoothly at these $ d $ double modes.
Hence, the theory is well behaved and free of massless doublers at the phase
boundary $ m_0 = 2 r_w $ in the free fermion limit but becomes topologically
trivial in any background gauge field. This "anomalous" phenomenon seems to
be of nonperturbative or topological origin since it cannot be understood
from the viewpoint of perturbation theory. Further discussions of the
behaviors around the phase boundaries are given in the Appendix.

For $ 2 r_w < m_0 \le 4 r_w $, there are $ d $ doubled modes each having
only one nonzero ( $ \pi/a $ ) momentum component which gives
$ u(p) < 0 $ and $ T(p) \ne 0 $, hence they are restored to the theory
and each one of them carries negative chiral charge. When we turn
on the background gauge field with topological charge $Q_{top}$,
these $ d $ double modes produce an index $ - d Q_{top} $ which adds to the
index $ +Q_{top} $ of the primary mode, thus gives the total index
$ (1-d) Q_{top} $. It agrees with numerical experiments for $ D $ in smooth
background gauge fields except at the phase boundary, $ m_0 = 4 r_w $,
where the actual index is always zero.

For $ 4 r_w < m_0 \le 6 r_w $ ( provided that $ d \ge 4 $ ), there
are additional $ d(d-1)/2 $ doubled modes restored to the theory.
Each of them has only two nonzero momentum components which
give $ u(p) < 0 $ and $ T(p) \ne 0 $, hence carries positive
chiral charge. Then the total chiral charge of the theory is
$ 1 - d + d(d-1)/2 $ after taking into account of the primary mode,
the $ d $ doubled modes having one non-zero momentum component,
and $ d(d-1)/2 $ doubled modes having two non-zero momentum components.
When we turn on the background gauge field with topological
charge $ Q_{top} $, the index is equal to $ ( 2 - 3 d + d^2 ) Q_{top} / 2 $.
This is a topologically improper phase.

The process of dividing $ m_0 $ into topologically
improper phases each of width $ 2 r_w $ continues until $ m_0 = 2 d r_w $,
then for $ m_0 > 2 d r_w $, $ T(p) \ne 0 $ for all doubled modes and they
are all restored to the theory. The index must be zero for any gauge
configurations, thus the theory is topologically trivial for
$ m_0 > 2 d r_w $.

Now we can pull the exact solutions of eigenvalues and the fermion
propagator together to obtain a complete picture of the topological phases
in the free fermion limit. From the exact solutions of eigenvalues,
Eq. (\ref{eq:eigen}), and the fermion propagator,
Eqs. (\ref{eq:SF0}), (\ref{eq:Tp}) and (\ref{eq:TpBZ}),
we immediately obtain that {\em the $+2$ eigenmodes must have $ T(p)=0 $ in
the fermion propagator, thus are completely decoupled from the theory;
while a zero mode has $ T(p)=| u(p) | $, hence its presence depends on
the value of $ m_0 $}. Therefore, using  Eq. (\ref{eq:indQ5}),
from the total chiral charge of the zero modes, we can obtain a formula
for the index of $ D $ in a smooth backgroundgauge field.
From Eqs. (\ref{eq:TpBZ})-(\ref{eq:uBZ}), we obtain the general
formula for the total chiral charge of the zero modes in each phase,
\bea
\label{eq:Q5}
Q_5^{(k)} = \sum_{i=1}^{k}
\frac{ (-1)^{i-1} d \ ! }{ (i-1) \ ! \ (d - i + 1) \ ! } \
= \frac{ (-1)^{k+1} d \ ! } { d \ \Gamma ( 1 + d - k ) \ \Gamma ( k ) },
\hspace{4mm} k=1, \cdots, d.
\eea
where the integer $ k $ denotes the $k$-th phase region,
$ 2(k-1) r_w  < m_0 \le 2k r_w $ which includes the upper phase boundary,
and $ Q_5 = 0 $ for $ m_0 \le 0 $ and $ m_0 > 2 d r_w $.
Inserting (\ref{eq:Q5}) into Eq. (\ref{eq:indQ5}), we obtain the
index of $ D $ in each phase as the following
\beq
\label{eq:index_pert}
\mbox{index}[D^{(k)}] = \left\{  \begin{array}{ll}
 0    &   \mbox{ if $ k = 0 $ \ for \ $ m_0 \le 0  $,  }      \\
\frac{ (-1)^{k+1} d \ ! }{ d \ \Gamma (1+d-k) \ \Gamma(k)  } \ Q_{top}
      &   \mbox{ if  $ k=1,\cdots,d $ \ for \ }            \\
$ $   &   \mbox{ \ \ $ 2(k-1) r_w < m_0 \le 2k r_w  $,   }  \\
 0    &   \mbox{ if $ k=d+1 $ \ for \ $ m_0 > 2dr_w $.   }  \\
                  \end{array}
              \right.
\eeq
%
It is evident that (\ref{eq:index_pert}) satisfies the exact
reflection symmetry $ \mbox{index}[D^{(k)}] = -\mbox{index}[D^{(d+1-k)}] $
( equivalent to Eq. (\ref{eq:ref_sym_1}) )
except at each one of the phase boundaries ( $ m_0 = 2 k r_w, k=1,\cdots,d $ )
where the index is nonzero and equal to the index on the LHS of the boundary.
Then it is obvious that (\ref{eq:index_pert}) is invalid at the symmetry
axis $ m_0 = d r_w $ where the index must be zero,
Eq. (\ref{eq:sym_axis}), according to the exact reflection symmetry.
In general, we can prove that (\ref{eq:index_pert}) breaks down at each
one of the phase boundaries ( $ m_0 = 2 k r_w, k=1,\cdots,d $ ). \newline
{\it Proof : } \newline
Assuming Eq. (\ref{eq:index_pert}) holds at the phase boundary
$ m_0 = x $, then we have
\bea
\label{eq:pf1}
\mbox{index}[D( x )] =  \mbox{index}[D( x - \epsilon )]
\eea
where $ \epsilon $ is an infinitesimal positive real number.
Then, the image of $ m_0 = x $ with respect to the mirror
at $ m_0 = d r_w  $ is $ m_0^{'} = 2dr_w - x $, and its index is
\bea
\label{eq:pf2}
\mbox{index} \left[ D( 2dr_w - x ) \right] = -\mbox{index}[D( x )]
\eea
according to the exact reflection symmetry, Eq. (\ref{eq:ref_sym_1}).
Now consider the image of $ m_0 = x - \epsilon $, and its index is
\bea
\label{eq:pf3}
\mbox{index} \left[ D( 2dr_w - x +\epsilon) \right]
= -\mbox{index}[D( x - \epsilon )]
\eea
Then from Eqs. (\ref{eq:pf1})-(\ref{eq:pf3}), we obtain
\bea
\label{eq:pf4}
  \mbox{index} \left[ D( 2dr_w - x + \epsilon) \right]
= \mbox{index} \left[ D( 2dr_w - x ) \right]
\eea
This leads to the contradiction since the indices of two neighbouring phases
must be different. Therefore Eq. (\ref{eq:pf1}) must be invalid
since the exact reflection symmetry is always true under any
circumstances. This completes the proof that (\ref{eq:index_pert})
fails at each one of the phase boundaries
( $ m_0 = 2 k r_w, k=1,\cdots,d $ ).

In Table 1, we list the real eigenmodes of $ D $ in the free fermion
limit on a two dimensional lattice with periodic boundary conditions.
The lattice can be the infinite lattice or a finite lattice with
even number of sites in each dimension. First, we recall that the zero
modes and $+2$ eigenmodes of $ D $ are both chiral \cite{twc98:4},
thus their chiral charges can be determined
by the number of nonzero ( $\pi/a$ ) momentum components of this
fermion mode.  We also recall that for each real eigenvalue
of $ D $ in two dimensions, it has degeneracy 2 which consists of a pair of
eigenmodes of $ \pm 1 $ chiralities, while complex eigenvalues must come in
complex conjugate pairs with chirality zero. The chirality degeneracy of
real eigenmodes is not shown explicitly in Table 1.
In the first phase, $ -\infty < m_0 \le 0 $, there is no zero
modes and the theory in fact does not describe a massless fermion.
The factor $ T(p) $ is zero at the origin $ p = 0 $ as well as at the
corners of the Brillouin zone.
In the second phase, $ 0 < m_0 < 2 r_w $, there is one zero mode
with positive chiral charge. The theory in this phase correctly describes
a single flavor of massless fermion.
At the phase boundary, $ m_0 = 2 r_w $, two $ +2 $ real
eigenmodes and their chirality partners become two pairs of complex
conjugate eigenmodes ( $ 1 \pm i $ ) which are decoupled from the theory,
i.e., $ T(p) = 0 $. The number of zero modes remains the same.
These two pairs of complex conjuagte eigenmodes become two zero modes
( in chirality pairs ) in the third phase, $  2 r_w < m_0 < 4 r_w $.
At the phase boundary, $ m_0 = 4 r_w $, the number of zero modes remains
the same, but the $ +2 $ eigenmode ( and its chirality partner ) become a
complex conjugate pair ( $ 1 \pm i $ ). Then they reappear as a zero mode
( in chirality pair ) in the last phase, $ 4 r_w < m_0 < \infty $, in which
all real eigenmodes are zero modes. We notice the interplay between the zero
modes and the $ +2 $ eigenmodes when we change the value of $ m_0 $ from
$-\infty$ to $+\infty$. Only in the region $ 0 < m_0 < 2 r_w $ and the
critical point $ m_0 = 2 r_w $, the theory has the correct particle contents
to describe a single flavor of massless fermion, while other regions
suffer from either species doubling or absence of zero modes. The total
chiral charge of zero modes in each phase is listed in the last
column, from which we can predict the index of $ D $ in a smooth
background gauge field according to Eq. (\ref{eq:indQ5}), i.e.,
Eq. (\ref{eq:index_pert}).
As we will see, when we turn on a smooth background gauge field
with non-zero topological charge,
Eq. (\ref{eq:index_pert}) is satisfied exactly for all phases except at the
phase boundaries where the index is always zero regardless of the total
chiral charge ( see Table 3 ).

In Table 2, we list the real eigenmodes of $ D $ in the free fermion limit on
a four dimensional lattice with periodic boundary conditions. Although it
has many more phases than Table 1 for two dimensions,
however, it shares the same essential features of Table 1, namely,
only the region $ 0 < m_0 < 2 r_w $ and the critical point
$ m_0 = 2 r_w $ can describe a single flavor of massless fermion.
We also recall that for each real eigenvalue of $ D $ in four dimensions,
it has degeneracy 4 which consists of two pairs of eigenmodes of $ \pm 1 $
chiralities, while complex eigenvalues come in complex conjugate pairs
with chirality zero and with degeneracy 2.
The chirality degeneracy of real eigenmodes is not shown explicitly
in Table 2. We will not go through each phase in Table 2 as what we have
done for Table 1 since it is straightforward to obtain these results
from the exact solutions of eigenvalues, Eq. (\ref{eq:eigen}) and the
fermion propagator, Eqs. (\ref{eq:SF0}), (\ref{eq:Tp}) and (\ref{eq:TpBZ}).

Our next task is to turn on the background gauge field to see to what
extent the formula, Eq. (\ref{eq:index_pert}), which relies on the weak
coupling perturbation theory and the exact solutions in the free fermion
limit, is satisfied in a background gauge field with nonzero topological
charge, as well as to see the displacements of the phase boundaries with
respect to the background gauge field.

{\footnotesize
\begin{table}
\begin{center}
\begin{tabular}{|c|c|c|c|c|c|c|}
\hline
 & real eigenvalues & multiplicity & $ \gamma_5^{'} $ & total chiral charge \\
 &                  &              &                  & of zero modes       \\
\hline
\hline
$ -\infty < m_0 \le 0 $  &   2  &   2  &  $+\gamma_5$  &        \\
                         &   2  &   2  &  $-\gamma_5$  &   0    \\
\hline
$ 0 < m_0 < 2 r_w $      &   2  &   1  &  $+\gamma_5$  &        \\
                         &   2  &   2  &  $-\gamma_5$  &        \\
                         &   0  &   1  &  $+\gamma_5$  &   1    \\
\hline
$  m_0 = 2 r_w $         &   2  &   1  &  $+\gamma_5$  &        \\
                         &   0  &   1  &  $+\gamma_5$  &   1    \\
\hline
$ 2 r_w < m_0 < 4 r_w $  &   2  &   1  &  $+\gamma_5$  &        \\
                         &   0  &   1  &  $+\gamma_5$  &        \\
                         &   0  &   2  &  $-\gamma_5$  &  -1    \\
\hline
$  m_0 = 4 r_w $         &   0  &   1  &  $+\gamma_5$  &        \\
                         &   0  &   2  &  $-\gamma_5$  &  -1    \\
\hline
$ 4 r_w < m_0 < \infty $ &   0  &   2  &  $+\gamma_5$  &        \\
                         &   0  &   2  &  $-\gamma_5$  &   0    \\
\hline
\end{tabular}
\end{center}
\caption{The real eigenmodes of $ D $ in the free fermion limit on a two
dimensional lattice with periodic boundary conditions.
The first column lists the ranges of $ m_0 $ values, the second column the
real eigenvalues, the third column the multiplicities of the eigenvalues in
the momentum space, the fourth column the effective $ \gamma_5 $ of the
eigenmode where the sign of $ \gamma_5 $ is used to denote the chiral
charge of the eigenmode, and the last column is the total chiral charge
of the zero modes. These results are obtained analytically from the
exact solutions on a finite lattice of even number of sites in each
dimension, or the infinite lattice.}
\label{table:1}
\vskip 0.2 true in
\end{table}
}
 
{\footnotesize
\begin{table}
\begin{center}
\begin{tabular}{|c|c|c|c|c|c|c|}
\hline
 & real eigenvalues & multiplicity & $ \gamma_5^{'} $ & total chiral charge \\
 &                  &              &                  & of zero modes       \\
\hline
\hline
$ -\infty < m_0 \le 0 $  &   2  &  8  &  $+\gamma_5$ &      \\
                         &   2  &  8  &  $-\gamma_5$ &  0   \\
\hline
$ 0 < m_0 < 2 r_w $      &   2  &  7  &  $+\gamma_5$ &      \\
                         &   2  &  8  &  $-\gamma_5$ &      \\
                         &   0  &  1  &  $+\gamma_5$ &  1   \\
\hline
$  m_0 = 2 r_w $         &   2  &  7  &  $+\gamma_5$ &      \\
                         &   2  &  4  &  $-\gamma_5$ &      \\
                         &   0  &  1  &  $+\gamma_5$ &  1   \\
\hline
$ 2 r_w < m_0 < 4 r_w $  &   2  &  7  &  $+\gamma_5$ &      \\
                         &   2  &  4  &  $-\gamma_5$ &      \\
                         &   0  &  1  &  $+\gamma_5$ &      \\
                         &   0  &  4  &  $-\gamma_5$ & -3   \\
\hline
$  m_0 = 4 r_w $         &   2  &  1  &  $+\gamma_5$ &      \\
                         &   2  &  4  &  $-\gamma_5$ &      \\
                         &   0  &  1  &  $+\gamma_5$ &      \\
                         &   0  &  4  &  $-\gamma_5$ & -3   \\
\hline
$ 4 r_w < m_0 < 6 r_w $  &   2  &  1  &  $+\gamma_5$ &      \\
                         &   2  &  4  &  $-\gamma_5$ &      \\
                         &   0  &  7  &  $+\gamma_5$ &      \\
                         &   0  &  4  &  $-\gamma_5$ &  3   \\
\hline
$  m_0 = 6 r_w $         &   2  &  1  &  $+\gamma_5$ &      \\
                         &   0  &  7  &  $+\gamma_5$ &      \\
                         &   0  &  4  &  $-\gamma_5$ &  3   \\
\hline
$ 6 r_w < m_0 < 8 r_w $  &   2  &  1  &  $+\gamma_5$ &      \\
                         &   0  &  7  &  $+\gamma_5$ &      \\
                         &   0  &  8  &  $-\gamma_5$ & -1   \\
\hline
$  m_0 = 8 r_w $         &   0  &  7  &  $+\gamma_5$ &      \\
                         &   0  &  8  &  $-\gamma_5$ & -1   \\
\hline
$ 8 r_w < m_0 < \infty $ &   0  &  8  &  $+\gamma_5$ &      \\
                         &   0  &  8  &  $+\gamma_5$ &  0   \\
\hline
\end{tabular}
\end{center}
\caption{The real eigenmodes of $ D $ in the free fermion limit on a four
dimensional lattice with periodic boundary conditions.
The first column lists the ranges of $ m_0 $ values, the second column the
real eigenvalues, the third column the multiplicities of the eigenvalues
in the momentum space, the fourth column the effective $ \gamma_5 $ of the
eigenmode where the sign of $ \gamma_5 $ is used to denote the chiral
charge of the eigenmode, and the last column the total chiral charge
of the zero modes. These results are obtained analytically from the
exact solutions on a finite lattice of even number of sites in each
dimension, or the infinite lattice.}
\label{table:2}
\end{table}
}

\section{Topological phase diagrams \\
         for d=2 and d=4 }

In this section we turn on the background gauge field to investigate
topological phases of $ D $ on two dimensional and four dimensional
lattices respectively. First, we define the background gauge fields
in the following.
On a 4-dimensional torus ( $ x_{\mu} \in [0,L_{\mu}], \mu = 1, \cdots, 4 $ ),
the simplest nontrivial $ SU(2) $ gauge fields can be represented as
\bea
\label{eq:A1}
A_1(x) &=& \tau \left[   \frac{ 2 \pi h_1 }{L_1}
                       - \frac{ 2 \pi q_1 x_2 }{ L_1 L_2 }
      +  A_1^{(0)} \sin \left( \frac{ 2 \pi n_2 }{L_2} x_2 \right) \right] \\
\label{eq:A2}
A_2(x) &=& \tau \left[ \frac{ 2 \pi h_2 }{L_2}
       +  A_2^{(0)} \sin \left( \frac{ 2 \pi n_1 }{L_1} x_1 \right) \right] \\
\label{eq:A3}
A_3(x) &=& \tau \left[  \frac{ 2 \pi h_3 }{L_3}
                      - \frac{ 2 \pi q_2 x_4 }{ L_3 L_4 }
       +  A_3^{(0)} \sin \left( \frac{ 2 \pi n_4 }{L_4} x_4 \right) \right] \\
\label{eq:A4}
A_4(x) &=& \tau \left[ \frac{ 2 \pi h_4 }{L_4}
        +  A_4^{(0)} \sin \left( \frac{ 2 \pi n_3 }{L_3} x_3 \right) \right]
\eea
where $ {\bf\tau} $ is one of the generators of $ SU(2) $ with the
normalization $ \tr( {\bf\tau}^2 ) = 2 $, and $ q_1 $ and $ q_2 $ are
integers. The global part is characterized by the topological charge
\beq
Q = \frac{1}{32\pi^2} \int d^4 x \ \epsilon_{\mu\nu\lambda\sigma} \
    \tr( F_{\mu\nu} F_{\lambda\sigma} ) = 2 q_1 q_2
\label{eq:ntop}
\eeq
which must be an integer. The harmonic parts are parameterized by four
real constants $ h_1 $, $ h_2 $, $ h_3 $ and $ h_4 $.
The local parts are chosen to be sinusoidal
fluctuations with amplitudes $ A_1^{(0)} $, $ A_2^{(0)} $, $ A_3^{(0)} $
and $ A_4^{(0)} $, and
frequencies $ \frac{ 2 \pi n_2 }{L_2} $, $ \frac{ 2 \pi n_1 }{L_1} $,
$ \frac{ 2 \pi n_4 }{L_4} $ and $ \frac{ 2 \pi n_3 }{L_3} $ where
$ n_1 $, $ n_2 $, $ n_3 $ and $ n_4 $  are integers.
The discontinuity of $ A_1(x) $ ( $ A_3(x) $ ) at $ x_2 = L_2 $
( $ x_4 = L_4 $ ) due to the global part
only amounts to a gauge transformation.
The field tensors $ F_{12} $ and $ F_{34} $ are continuous
on the torus, while other $ F's $ are zero.
To transcribe the continuum gauge fields to the lattice, we take the lattice
sites at $ x_\mu = 0, a, ..., ( N_\mu - 1 ) a $, where $ a $ is the lattice
spacing and $ L_\mu = N_\mu a $ is the lattice size.
Then the link variables are
\bea
\label{eq:U1}
U_1(x) &=& \exp \left[ \text{i} A_1(x) a \right] \\
\label{eq:U2}
U_2(x) &=& \exp \left[ \text{i} A_2(x) a
 + \text{i} \delta_{x_2,(N_2 - 1)a} \frac{ 2 \pi q_1 x_1 }{L_1} \tau \right] \\
\label{eq:U3}
U_3(x) &=& \exp \left[ \text{i} A_3(x) a \right] \\
\label{eq:U4}
U_4(x) &=& \exp \left[ \text{i} A_4(x) a
 + \text{i} \delta_{x_4,(N_4 - 1)a} \frac{ 2 \pi q_2 x_3 }{L_3} \tau \right]
\eea
The last term in the exponent of $ U_2(x) $ ( $ U_4(x) $ ) is included to
ensure that the field tensor $ F_{12} $ ( $ F_{34} $ ) which is defined by
the ordered product of link variables around a plaquette is continuous on
the torus. Similar construction can be carried out for the background gauge
field on a two dimensional lattice, which has been described in details in
\cite{twc98:4}, and we use the same notations as defined in Eqs. (7)-(11) in
\cite{twc98:4}.

{\footnotesize
\begin{table}
\begin{center}
\begin{tabular}{|c|c|c|c|c|c|}
\hline
                  & real eigenvalues & multiplicity & chirality & index \\
\hline
\hline
$ -\infty < m_0 \le 0.5472 $  &      &      &      &  0   \\
\hline
$ 0.5473 \le m_0 < 2.0 $      &   2  &   1  &  +1  &      \\
                              &   0  &   1  &  -1  &  1   \\
\hline
$  m_0 = 2.0 $                &      &      &      &  0   \\
\hline
$ 2.0 < m_0 \le 3.4527 $      &   2  &   1  &  -1  &      \\
                              &   0  &   1  &  +1  & -1   \\
\hline
$ 3.4528 \le m_0 < \infty $   &      &      &      &  0   \\
\hline
\end{tabular}
\end{center}
\caption{The index of $ D $ on a $ 6 \times 6 $ lattice in the background
gauge field with topological charge $ Q = 1 $, and other parameters
$ h_1 = 0.5 $, $ h_2 = 0.6 $, $ A_1^{(0)} = 0.7 $, $ A_2^{(0)} = 0.8 $
and $ n_1 = n_2 = 1 $. The Wilson parameter $ r_w $ is set to $ 1.0 $.
We note that the gauge field shifts the critical points from their values
in the free fermion limit. The reflection symmetry is satisfied exactly by
checking that the indices are odd under the reflection with respect to
the symmetry axis $ m_0 = 2.0 $, and the displacement at the first
phase boundary $ \delta_0 = 0.5472 $ is equal to minus of the displacement
$ \delta_2 = 3.4528 - 4.0 = -0.5472 $ at the third phase boundary,
and the location of the symmetry axis $ m_0 = 2.0 $ is invariant for
any background gauge field.
All critical points are accurate up to $ \pm 0.00005 $ except
the central critical point $ 2.0 $ is exact. The indices in the last column
agree exactly with Eq. (\ref{eq:index_pert})
except at the phase boundary ( $ m_0 = 2.0 $ ). }
\label{table:3}
\end{table}
}

\psfigure 5.0in -0.2in {fig:2d} {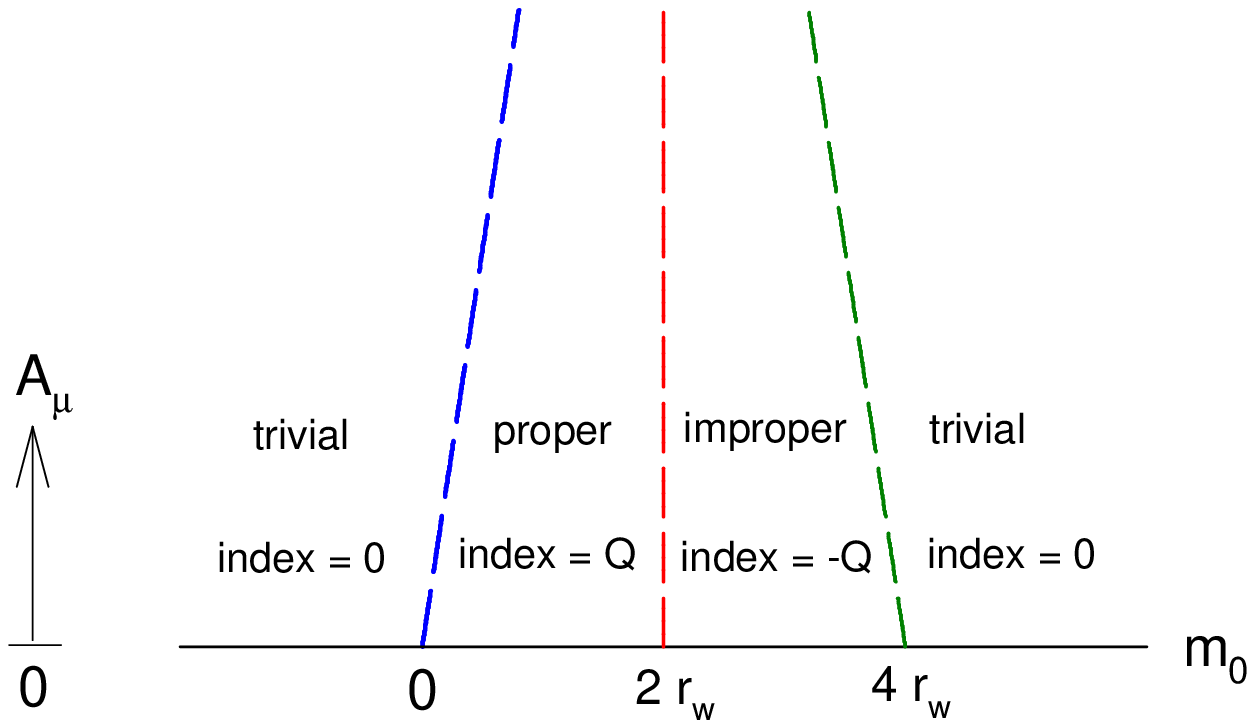} {
The topological phases of $ D $ on a two dimensional lattice
for smooth background gauge fields.
The phase boundaries denoted by dashed lines are sketched to indicate
their evolutions with respect to the background gauge field.
The reflection symmetry with respect to the
central critical line $ m_0 = 2 r_w $ is displayed explicitly.
Note that the central critical line $ m_0 = 2 r_w $ is invariant for
any background gauge configurations.
The phase transition points in the free fermion limit
( at the horizontal line ) are determined exactly. }

\psfigure 5.0in -0.2in {fig:4d} {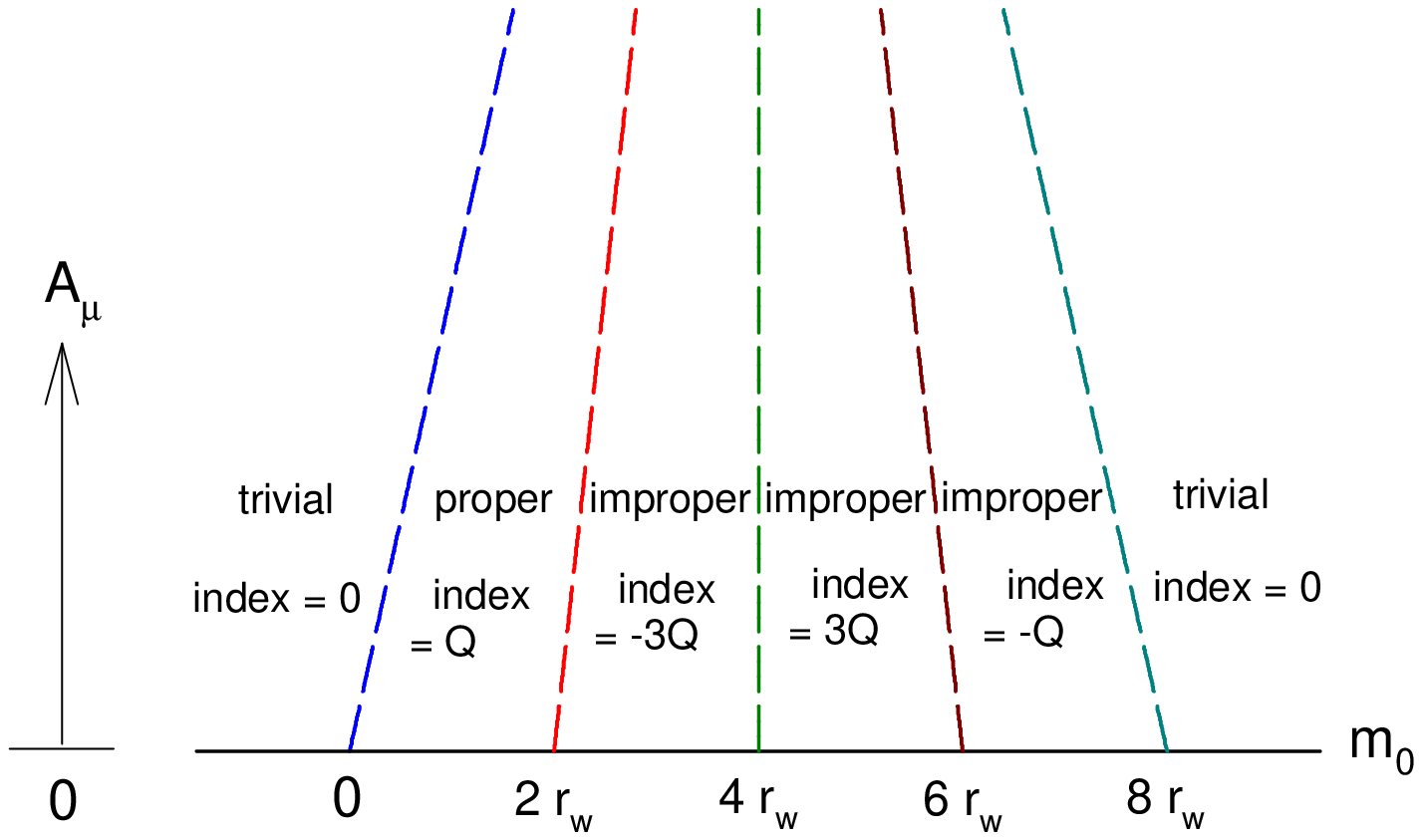} {
The topological phases of $ D $ on a four dimensional lattice
for smooth background gauge fields.
The phase boundaries with dashed lines are sketched to indicate
their evolutions with respect to the background gauge field.
The reflection symmetry with respect to
the central critical line $ m_0 = 4 r_w $ is shown explicitly.
Note that the central critical line $ m_0 = 4 r_w $ is invariant for
any background gauge configurations.
The critical points in the free fermion limit ( at the horizontal line )
are determined exactly. }

In Table 3, we list the real eigenmodes of $ D $ on a $ 6 \times 6 $
lattice in the background gauge field with topological charge $ Q = 1 $,
and other parameters $ h_1 = 0.5 $, $ h_2 = 0.6 $, $ A_1^{(0)} = 0.7 $,
$ A_2^{(0)} = 0.8 $ and $ n_1 = n_2 = 1 $. The Wilson parameter $ r_w $
has been set to 1.0. The indices of $ D $ in the last column are computed
by counting the zero modes according to $\mbox{index}(D) = n_{-} - n_{+}$.
They agree exactly with the formula, Eq. (\ref{eq:index_pert}),
with the total chiral charge of the zero modes obtained from the exact
solutions in the free fermion limit as listed in the last column of
Table 1, except at the phase boundaries where the index is always zero
regardless of the total chiral charge.
The response of $ D $ in the first region is easily understood since it
has no zero modes in the free fermion limit and thus it cannot possess any
zero modes after we turn on the gauge field with nonzero topological charge.
The phase boundary has been shifted from $ m_0 = 0 $ to $ m_0 = 0.5472 $
due to the background gauge fields. The index of $ D $ is always zero in
this phase and $ D $ is called {\it topologically trivial.}
In the second phase, the index of $ D $ is always equal to the topological
charge of the gauge field, and $ D $ is called {\it topologically proper.}
Since $ D $ is free of species doubling and the chiral charge of the zero
modes is $ +1 $, thus $ D $ satisfies the Atiyah-Singer index theorem
exactly in this phase. However, at the phase boundary of the central
critical line $ m_0 = 2.0 $, $ D $ has the correct particle contents
in the free fermion limit, but it does not possess any zero modes after
turning on the topologically nontrivial background gauge field.
This cannot be understood from the viwepoint of perturbation theory,
though we know that the index must be zero at the symmetry
axis, Eq. (\ref{eq:sym_axis}), due to the exact reflection symmetry.
Further discussions of the behaviors at the phase boundary are
given in the Appendix.
%
%
If we step into the next phase, we find that the index is opposite to the
topological charge, in complete agreement with the
formula (\ref{eq:index_pert}) with the total
chiral charge $ Q_5 = -1 $ in the free fermion limit, as listed in Table 1.
In the last phase, $ D $ has four species of massless fermion modes
in the free fermion limit, similar to the case of the naive lattice fermion,
then there is no zero modes and the index is zero for any background gauge
field. We note that the displacements of the phase boundaries
due to the background gauge field must obey the exact reflection
symmetry $ \delta_k = - \delta_{d-k}, k=0, \cdots, d $,
which is an immediate consequence of Eq. (\ref{eq:ref_sym_1}).
In Table 3, $ \delta_0 = 0.5472 $ at the first phase boundary is equal to
minus of the displacement $ \delta_2 = 3.4572 - 4.0 = - 0.5472 $ at the
third phase boundary. The central phase boundary $ m_0 = 2.0 $ is
invariant for any background gauge field.
In Fig. 2, we sketch the topological phase diagram of $ D $ in smooth
background gauge fields on a two dimensional lattice, which summarizes
the gross features of a lot of numerical experiments. The discrete
symmetry, $ \mbox{index}[ D( 4 r_w - m_0 ) ] = - \mbox{index} [ D( m_0 )] $
is satisfied exactly in all cases.
The index in each phase agrees exactly with the formula
(\ref{eq:index_pert}), with the total chiral charge of the zero modes
obtained from the exact solutions in the free fermion limit as listed in
the last column of Table 1, except at the phase boundaries
where the index is always zero regardless of the total chiral charge.

{\footnotesize
\begin{table}
\begin{center}
\begin{tabular}{|c|c|c|c|c|c|}
\hline
                  & real eigenvalues & multiplicity & chirality & index \\
\hline
\hline
$ -\infty < m_0  \le 0.3762 $              &      &      &      &  0   \\
\hline
$ 0.3763 \le m_0 \le 2.1880 $              &   2  &   2  &  +1  &      \\
                                           &   0  &   2  &  -1  &  2   \\
\hline
$  m_0 = 2.1881    $                       &      &      &      &  0   \\
\hline
$  2.1882 \le m_0 < 4.0  $                 &   2  &   6  &  -1  &      \\
                                           &   0  &   6  &  +1  & -6   \\
\hline
$  m_0 = 4.0 $                             &      &      &      &  0   \\
\hline
$ 4.0 < m_0 \le 5.8118 $                   &   2  &   6  &  +1  &      \\
                                           &   0  &   6  &  -1  &  6   \\
\hline
$  m_0 = 5.8119 $                          &      &      &      &  0   \\
\hline
$ 5.8120 \le m_0 \le 7.6237  $             &   2  &   2  &  +1  &      \\
                                           &   0  &   2  &  -1  & -2   \\
\hline
$ 7.6238 \le m_0 < \infty    $             &      &      &      &  0   \\
\hline
\end{tabular}
\end{center}
\caption{The index of $ D $ on a $ 4^4 $ lattice in the background
gauge field with topological charge $ Q = 2 $, and other parameters
$ h_1 = 0.1 $, $ h_2 = 0.2 $, $ h_3 = 0.3 $, $ h_4 = 0.4 $,
$ A_1^{(0)} = A_2^{(0)} = A_3^{(0)} = A_4^{(0)} = 0 $.
The reflection symmetry,
$ \mbox{index}[ D( 8 - m_0 ) ] = - \mbox{index} [ D( m_0 )] $
is satisfied exactly, in particular,
the displacements of the phase boundaries due to the background
gauge field satisfy the relationship
$ \delta_k = -\delta_{4-k}, k = 0, \cdots, 4 $.
All critical points are accurate up to $ \pm 0.00005 $ except the
central critical point $ 4.0 $ is exact.
The indices in the last column
agree exactly with the formula (\ref{eq:index_pert}),
except at the phase boundaries $ m_0 = 2.1881, \ 4.0, \ 5.8119 $.
}
\label{table:4}
\end{table}
}

In Table 4, we list the real eigenmodes of $ D $ on a $ 4^4 $
lattice in the background gauge field with topological charge $ Q = 2 $,
and other parameters $h_1 = 0.1$, $h_2 = 0.2$, $h_3 = 0.3$, $h_4 = 0.4$,
$ A_1^{(0)} = A_2^{(0)} = A_3^{(0)} = A_4^{(0)} = 0 $.
The Wilson parameter $ r_w $ is set to $ 1.0 $.
There are more phases regions than those in two dimensions
( see Table 3 ), and the locations of all critical points are shifted
by the background gauge field except the central critical point $ m_0 = 4.0 $.
It is evident that the reflection symmetry, Eq. (\ref{eq:ref_sym_1}) is
satisfied exactly. We note that the displacements of the phase boundaries
due to the background gauge field are
$ \delta_0 = 0.3762 $, $ \delta_1 = 2.1881 - 2.0 = 0.1881 $,
$ \delta_2 = 0 $, $ \delta_3 = 5.8119 - 6.0 = -0.1881 $ and
$ \delta_4 = 7.6238 - 8.0 = -0.3762 $, which satisfy
the exact reflection symmetry
$ \delta_k = - \delta_{4-k}, k=0, \cdots, 4 $.
We do not go through the detailed descriptions for each
phase as we have done for Table 3.
%
%
We have also performed extensive numerical tests in the topologically proper
phase as well as in the improper phases. For smooth background gauge fields
with nonzero topological charge, the exact zero modes with definite chirality
are reproduced to a very high precision on a finite four dimensional
lattice, and the index theorem is satisfied exactly in the topologically
proper phase. The zero modes are also very stable under random fluctuations
as defined in Eq. (43) of ref. \cite{twc98:4}.
This completes our intended investigations for $ D $ in four
dimensional smooth background gauge fields as outlined in \cite{twc98:4}.
In Fig. 3, we sketch the topological phase diagram of $ D $ in smooth
background gauge fields on a four dimensional lattice with even number of
sites in each dimension, which summarizes the gross features of a lot of
numerical experiments. The discrete symmetry,
$ \mbox{index}[ D( 8 r_w - m_0 ) ] = - \mbox{index} [ D( m_0 )] $
is satisfied exactly in all cases.
The index in each phase agrees exactly with the formula,
Eq. (\ref{eq:index_pert}), except at the phase boundaries
where the index is always zero regardless of the total chiral charge.

It is interesting to note that in ref. \cite{ehn98:1}, the spectral flow
of the hermitian Wilson-Dirac operator $ H_w(m_0) = \gamma_5 D_w( m_0 ) $
is studied numerically for smooth $ SU(2) $ instanton backgrounds on $ 8^4 $
lattice, and the symmetry $ -H_w( 8 - m_0, U ) = H_w( m_0, -U ) $ is displayed.
Since the index of $ D $ can also be computed by studying the level crossings
in the spectral flow of $ H_w $ as a function of $ m_0 $
[ see Eq. (\ref{eq:index_D} ) in the Appendix ], the results of
ref. \cite{ehn98:1} provide another four-dimensional verification
of the reflection symmetry (\ref{eq:ref_sym_1}) proved in Section 2.
Furthermore, in ref. \cite{ehn98:7}, the spectrum of the hermitian
Neuberger-Dirac operator $ H = \gamma_5 D $ in a smooth $ SU(2) $
instanton background is investigated in the phase $ 0 < m_0 < 2 $.
The chiral properties of $ D $ ( $ H $ ) are verified and the pole
method \cite{hn98:6} appears to be the most cost effective implementation.

\section{ Summary and Discussions }

{\footnotesize
\begin{table}
\begin{center}
\begin{tabular}{|c|c|c|c|c|c|}
\hline
                  & real eigenvalues & multiplicity & chirality & index \\
\hline
\hline
$ -\infty < m_0 \le 0.7002 $      &      &      &      &  0   \\
\hline
$ 0.7003 \le m_0 \le 0.9050 $     &   2  &   1  &  +1  &      \\
                                  &   0  &   1  &  -1  &  1   \\
\hline
$ 0.9051 \le m_0 \le 0.9662 $     &      &      &      &  0   \\
\hline
$ 0.9663 \le m_0 \le 1.4545 $     &   2  &   1  &  -1  &      \\
                                  &   0  &   1  &  +1  & -1   \\
\hline
$ 1.4546 \le m_0 \le 1.5592 $     &      &      &      &  0   \\
\hline
$ 1.5593 \le m_0 < 2.0 $          &   2  &   1  &  -1  &      \\
                                  &   0  &   1  &  +1  & -1   \\
\hline
$  m_0 = 2.0 $                    &      &      &      &  0   \\
\hline
$ 2.0 < m_0 \le 2.4407 $          &   2  &   1  &  +1  &      \\
                                  &   0  &   1  &  -1  & +1   \\
\hline
$ 2.4408 \le m_0 \le 2.5454 $     &      &      &      &  0   \\
\hline
$ 2.5455 \le m_0 \le 3.0337 $     &   2  &   1  &  +1  &      \\
                                  &   0  &   1  &  -1  & +1   \\
\hline
$ 3.0338 \le m_0 \le 3.0949 $     &      &      &      &  0   \\
\hline
$ 3.0950 \le m_0 \le 3.2997 $     &   2  &   1  &  -1  &      \\
                                  &   0  &   1  &  +1  & -1   \\
\hline
$ 3.2998 \le m_0 < \infty $       &      &      &      &  0   \\
\hline
\end{tabular}
\end{center}
\caption{The index of $ D $ on a $ 6 \times 6 $ lattice in the background
gauge field with topological charge $ Q = 1 $, and other parameters
$ h_1 = 0.5 $, $ h_2 = 0.6 $, $ A_1^{(0)} = 2.0 $, $ A_2^{(0)} = 0.8 $
and $ n_1 = n_2 = 1 $. The Wilson parameter $ r_w $ is set to $ 1.0 $.
The topological phases have bifurcated from 4 into 13 regions in two
steps. The reflection symmetry with respect to the central critical line
$ m_0 = 2.0 $ is satisfied exactly by the indices in all phases as well as
the positions of the critical points.
All critical points are accurate up to $ \pm 0.00005 $ except the
central critical point $ 2.0 $ is exact.}
\label{table:5}
\end{table}
}

We have investigated the topological phase diagrams of the
Neuberger-Dirac operator $ D $ with respect to the negative mass parameter
$ m_0 $ and the background gauge field. When the gauge configurations
are far from being rough, the phase diagrams can be schematically
represented in Figs. \ref{fig:dd}, \ref{fig:2d} and \ref{fig:4d} for
$d$-dimensions, $2$-dimensions and $4$-dimensions respectively.
Each topological phase diagram ( Figs. 1-3 ) possesses the exact
reflection symmetry (\ref{eq:ref_sym_1}). The symmetry axis
$ m_0 = d r_w $ is a phase boundary with zero index
and its position is invariant for any background gauge field.
Among the many different phases in each phase diagram, there is only one
phase in which the Atiyah-Singer index theorem is satisfied exactly, while
other phases are either trivial or improper.

The formula (\ref{eq:index_pert}) for the index of $ D $ in each
phase is derived by obtaining the total chiral charge of the zero
modes in the exact solution of the free fermion propagator.
For smooth gauge configurations, (\ref{eq:index_pert}) is verified
to be correct by numerical experiments on two and four dimensional
finite lattices with even number of sites in each direction,
except at the phase boundaries starting at
( $ m_0 = 2 k r_w, k=1,\cdots,d $ ) where the index is always zero for
any background gauge fields. Moreover, using the exact reflection symmetry,
we prove that (\ref{eq:index_pert}) must break down at the phase boundaries.
This indicates that there may be some nonperturbative or
topological effects happening at the phase boundaries. Further discussions
of the behaviors at the phase boundaries are given in the Appendix.
Now it is natural to generalize (\ref{eq:index_pert}) by
incorporating the displacements of the phase boundaries due to the
background gauge field, and also imposing the index zero at all phase
boundaries, then (\ref{eq:index_pert}) becomes
\beq
\label{eq:index_pert_sym}
\mbox{index}[D^{(k)}] = \left\{  \begin{array}{ll}
\frac{ (-1)^{k+1} d \ ! }{ d \ \Gamma (1+d-k) \ \Gamma(k)  } \ Q_{top}
     & \mbox{if \ $ 2(k-1) r_w + \delta_{k-1} < m_0 < 2 k r_w + \delta_{k} $} \\
$ $  & \mbox{ \ \  \ for $ k=1,\cdots,d $ ; }            \\
 0   & \mbox{ if \ $ m_0 = 2 k r_w + \delta_k $ for $ k=0,\cdots,d $.   }  \\
                  \end{array}
              \right.
\eeq
where $ \delta_k, k=0, \cdots, d $ are the displacements of the phase
boundaries due to the background gauge field, which satisfy
\bea
\label{eq:delta_sym}
\delta_k = - \delta_{d - k}.
\eea
Equations (\ref{eq:index_pert_sym}) and (\ref{eq:delta_sym})
constitute the general formula for the index of Neuberger operator
in a smooth background gauge field, as a function of the negative mass
parameter $ m_0 $. The exact reflection symmetry (\ref{eq:ref_sym_1})
is obviously satisfied by Eqs. (\ref{eq:index_pert_sym}) and
(\ref{eq:delta_sym}).

It is tempting to go beyond the limit of smooth background gauge field by
increasing the field strength or its roughness, in order to explore what
lies ahead in the upper portions of these phase diagrams.
However, we quickly run into difficulties as
{\it the phase boundaries bifurcate into many different phases, reminiscent
the chaos in dynamical systems.}  Even for $ D $ in a topologically proper
phase, an infinitesimal change in $ m_0 $ may end up having $ D $ in a
trivial or improper phase. For a given rough gauge
configuration, we do not know {\it a priori } what is the range of
$ m_0 $ values for $ D $ to be in the topologically proper phase.
We can easily demonstrate the bifurcation of the topologcial
phases of $ D $ by increasing the amplitudes of the local sinusoidal
fluctuations of the background gauge field on a two dimensional lattice.
In Table 5, all parameters are the same as Table 3 except the amplitude
$ A_1^{(0)} $ is increased from $ 0.7 $ to $ 2.0 $. We see that there
are total 13 phase regions which are resulted from bifurcations in two steps.
It is evident that formula (\ref{eq:index_pert_sym}) fails completely.
However, the exact reflection symmetry, Eq. (\ref{eq:ref_sym_1}), for the
index in each phase as well as the positions of the phase boundaries,
is still satisfied exactly, though we cannot predict the number of
different topological phases and the index of $ D $ in each phase.
We do not know precisely the conditions under which the bifurcations
may happen, but one thing for sure is that the index is identically
zero at all phase boundaries.
%
%

We have also measured the indices of $ D $ with $ m_0 $ fixed at $ 1.0 $
for an ensemble of gauge configurations from quenched simulations, and
compared them to the geometrically defined topological charges of
the gauge configurations. It turns out that the index does not always
agree with the geometrically defined topological charge.
The discrepancies may be due to the fact that the fixed value $ m_0 = 1.0 $
does not fall into the topologically proper phase for all gauge
configurations. If that is the case, we can proceed in the
following way : for a given gauge configuration,
we measure the index of $ D(m_0) $ for sufficiently sampled $ m_0 $
values and exploit the reflection symmetry, then we should be able to
identify the correct index unambiguously. Otherwise, we must resolve
the discrepancies from the fundamental viewpoint, namely, the very
definition of topological charge on a lattice.

Finally we would like to remark that it is interesting to understand the
underlying dynamics which provoke {\em the collective motion} of the
eigenvalues of $ D $ ( or equivalently the {\em zero crossings}
of the spectral flow of the hermitian Wilson-Dirac operator
$ H_w = \gamma_5 D_w $ ) at the phase boundaries ( except for the first
and the last phase boundaries ) such that the index
always turns out to be zero, though the exact refection symmetry must impose
the zero index at the phase boundary $ m_0 = d r_w $.

\bigskip

\appendix

\section{ }

In this appendix, we discuss the spectral flow of the Neuberger operator
as a function of the negative mass parameter $ m_0 $,
in particular, its behaviors at the phase boundaries, in order to
gain a perspective on the underlying mechanism which gives the
zero index at the phase boundaries. Note that the index at the
phase boundary $ m_0 = d r_w $ ( the symmetry axis ) must be zero
due to the exact reflection symmetry (\ref{eq:ref_sym_1}).
Since the Neuberger operator can be
written in terms of the hermitian Wilson-Dirac operator
$ H_w(m_0) = \gamma_5 D_w(m_0) $,
\bea
D = \Id + V = \Id + \gamma_5 \frac{ H_w }{ \sqrt{ H_w^2 }  },
\eea
then the index of $ D $ is
\bea
\mbox{index}(D) = \frac{1}{2} \mbox{Tr} ( \gamma_5 V )
                = \frac{1}{2} \mbox{Tr} \left( \frac{H_w}{\sqrt{H_w^2}} \right)
                = \frac{1}{2} ( h_{+} - h_{-} )
\label{eq:index_D}
\eea
where Tr denotes the trace over the Dirac space and the position
space, $ h_{+} ( h_{-} ) $ is the number of positive ( negative )
eigenvalues of $ H_w $. So, the index of $ D $ can be obtained by
studying the level crossings of the low-lying eigenvalues of $ H_w $
\cite{ehn98:1}. However, the spectral flow of the Neuberger
operator $ D $ seems to be more complicated than that of the hermitian
Wilson-Dirac operator $ H_w $. Since eigenvalues of $ H_w(m_0) $ are
real, its spectral flow as a function of $ m_0 $ can be plotted as curves
on a plane. On the other hand, the eigenvalues of the Neuberger operator are
complex, lying in the unit circle with center at one on the complex
plane \cite{twc98:4}, therefore the spectral flow as a function of $ m_0 $
can be plotted as curves on the unit cylinder with
center at one in the three dimensional space formed by the complex plane
of the eigenvalues and the parameter $ m_0 $.
The relationship between the eigenvalues of $ D $ and $ H_w $ can be
written as
\bea
\label{eq:D_Hw}
\lambda = \Id + U^{\dagger} \gamma_5 U_H
                \frac{\lambda_H}{\sqrt{\lambda_H^2}} U_H^{\dagger} U
\eea
where $ \lambda $ ( $ \lambda_H $ ) is the diagonal matrix of eigenvalues
of $ D $ ( $ H_w $ ), $ U $ ( $ U_H $ ) is the unitary transformation matrix
formed by the eigenfunctions, satisfying
\bea
\label{eq:DS}
D &=& U \lambda U^{\dagger}  \\
\label{eq:HU}
H_w &=& U_H \lambda_H U_H^{\dagger}
\eea
It is evident that the spectral flow of $ D $ not only depends on
the spectral flow of $ H_w $, but also the overlap of their
eigenfunctions.

In general, due to the limited precisions of most numercial computations,
one cannot trace the spectral flow of $ D $ to any precisions one wishes.
However, even if one can trace the spectral flow to a very high precision,
one still do not understand why the eigenvalues behave in such a manner
at the phase boundaries so that the index must be zero, except for the
first and the last phase boundaries.
For example, let us describe the spectral flow of $ D $ around the
phase boundary $ m_0 = 2 $ in Table 3. Although we know that its
index must be zero due to the exact reflection symmetry, it is still
instructive to examine the spectral flow around this phase boundary.
At $ m_0 = 2 - \epsilon $, where $ \epsilon $ is any positive infinitesimal
number, the zero mode of negative chirality is accompanied by a $+2$
eigenmode of positive
chirality, and the rest of the eigenvalues are in complex conjugate pairs.
However, at $ m_0 = 2 $, the zero mode becomes a member of a complex
conjugate pair ( $ a_1 \pm i a_2 $ ), while the associated $+2$ eigenmode
becomes a member of another complex conjugate pair ( $ 2-a_3 \pm i a_4 $ ),
where $ a_i ( i=1, \cdots, 4 ) $ are positive small numbers.
Presumbaly these two pairs of complex conjugates have no direct correlations
since they are far apart. However, in order to create one additional complex
conjugate pair out of the zero mode and its associated $+2$ real eigenmode,
all other complex conjugate pairs must undergo a {\em collective motion}
in the unit circle such that all eigenvalues turn out to be in complex
conjugate pairs. Then at $ m_0 = 2 + \epsilon $, all complex conjugate
pairs again undergo another collective motion to yield a zero mode of
positive chirality and its associated $+2$ eigenmode of negative chirality.
It seems to be unlikely that such collective motions of the eigenvalues
at the phase boundaries can be described by any perturbation theory in
$ m_0 $ or the background gauge fields.

Now let us turn to the spectral flow of the hermitian Wilson-Dirac operator
$ H_w = \gamma_5 D_w $, which corresponds to the spectral flow of $ D $
described above. 
At $ m_0 = 2 - \epsilon $, the eigenvalues
of $ H_w $ are mostly in plus/minus pairs except two positive small
eigenvalues. Thus $ h_{+} - h_{-} = 2 $, and the index of $ D $ is one
according to Eq. (\ref{eq:index_D}). At $ m_0 = 2 $, one of the positive
small eigenvalues crosses zero\footnote{If one examines the crossing
numerically, one may find a seemingly discontinuity of order $ 10^{-8} $,
however, which is only due to the limited precision of one's numerical
computation.} and becomes a negative small eigenvalue.
The transition is exactly at the point $ m_0 = 2 $.
So $ h_{+} = h_{-} $, and the index is zero.
Then at $ m_0 = 2 + \epsilon $, the remaining positive small eigenvalue
also crosses zero and turns into a negative small eigenvalue.
Since $ h_{-} - h_{+} = 2 $, the index is $ -1 $.
Now the seemingly puzzle is why these two positive small
eigenvalues do not simultaneously become two negative small eigenvalues
at the point $ m_0 = 2 $, then the phase boundary of zero index
would not exist and the formula (\ref{eq:index_pert}) would also hold
for the phase boundaries.
However, this would lead to a contradiction to the exact reflection
symmetry (\ref{eq:sym_axis}) which must hold in general, and the only
way out is to have the spectral flow to produce zero index at the
phase boundary of the symmetry axis ( $ m_0 = d r_w $ ).
So, this puzzle seems to be resolved, but the underlying mechanism is
still not understood. In general, for phase boundaries other than the first
one ( stemming at $ m_0 = 0 $ ), the symmetry axis $ m_0 = d r_w $,
and the last one ( stemming at $ m_0 = 2 d r_w $ ),
we do not have any arguments to explain why there must be the number
$ \frac{1}{2} | h_{+} - h_{-} | $ of eigenvalues undergo a
sign-flipping transition when approaching one side of the phase boundary,
while the number $ \frac{1}{2} | h'_{+} - h'_{-} | $ the other side of
the phase boundary, such that the index must be zero at the phase boundary.
Furthermore, it seems to be more difficult to tackle what is the underlying
dynamics that provoke such transitions at the phase boundaries.

\bigskip
\bigskip
\bigskip

\vfill\eject
\flushpar
\noindent
{\bf Acknowledgement }
\bigskip

\noindent
I would like to thank Shailesh Chandrasekharan for drawing my attention
to a remark in his paper \cite{chand98:5}, while we were attending
Lattice 98 at Boulder. After this paper had posted to the archive hep-lat,
Chandrasekharan's results on the Schwinger model \cite{chand98:10} were also
posted. I also wish to thank Martin L\"uscher for his comment in a
correspondence. Finally I thank Herbert Neuberger for his
helpful discussion on the level crossings of the hermitian Wilson-Dirac
operator. This work was supported by the National Science Council,
R.O.C. under the grant number NSC88-2112-M002-016.

\bigskip
\bigskip
\bigskip

\flushpar

\end{document}